\documentclass{mn2e} 
\usepackage{epsfig}  
\usepackage{graphicx} 
\usepackage{amssymb, latexsym, amsopn}

\begin{document} 

 \title{The Great Attractor and the Shapley Concentration}

 \author[K. Bolejko and C. Hellaby]{
  Krzysztof Bolejko$^{1}$%
    \thanks{E-mail: bolejko@camk.edu.pl}
  and 
  Charles Hellaby$^{2}$%
    \thanks{E-mail: cwh@maths.uct.ac.za} \\
  $^1$Nicolaus Copernicus Astronomical Center,
      Polish Academy of Sciences, 
      ul. Bartycka 18,
      0-16 Warsaw,
      Poland \\
  $^2$Department of Mathematics and Applied Mathematics
      University of Cape Town
      Rondebosch,
      7701,
      Cape Town,
      South Africa
}

\date{\tt astro-ph/0604402}

\maketitle

\label{firstpage}

\begin{abstract}{
We develop a non-linear relativistic model of the Shapley Concentration (SC) and its environs, including the Great Attractor (GA) and the Local Group (LG).  We take the Shapley concentration as a major attractive centre, and we use the Lema\^{\i}tre-Tolman model.  We constrain our model with present day observations, plus the requirement that it have a physically reasonable evolution from small perturbations at last scattering.  We investigate possible mass and velocity distributions, and we find that the peculiar velocity maximum near the SC is $\sim 800$~km/s inwards, the density between GA and SC must be about $\sim 0.9$ times background, the mass of the GA is probably $4 - 6 \times 10^{15}$ M$_\odot$, the SC's contribution to the LG motion is negligible, and the value of the cosmological constant is not significant on this scale.
}
\end{abstract}
 
\begin{keywords}
cosmology: theory --- observations --- large--scale structure of Universe --- galaxies: clusters.
\end{keywords}

\section{Introduction}

The Great Attractor and the Shapley Concentration are the largest and most important cluster concentrations in the local Universe. The Great Attractor is mostly hidden behind our Milky Way (Woudt \& Kraan--Korteweg 2001). The first measurements of peculiar velocities of the galaxies in this region showed that the galaxies move towards the Great Attractor (Lynden--Bell et al. 1988). Although the motion towards the Great Attractor is significant, there is no evidence for any backside infall onto the Great Attractor. This suggests that the galaxy flow in this region is just a part of a larger flow, caused by some  more massive attractive center. Behind the Great Attractor there is the Shapley Concentration. The Shapley Concentration is not in the Zone of Avoidance and hence is more suitable for observations. 
 
\begin{figure} 
\includegraphics[scale=0.25]{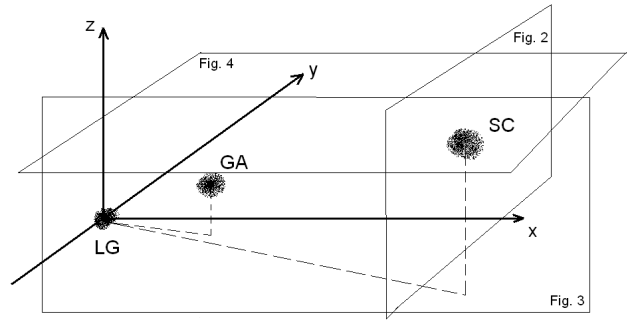} 
\caption{The schematic distribution of the Local Group (LG), the Great Attractor (GA) and the Shapley Concentration (SC) in space.  The coordinates $X,~Y,~Z$ are as described in the text.  The three surfaces correspond to Figs. 2-4}
\label{scm}
\end{figure}

\begin{figure} 
\includegraphics[scale=0.6]{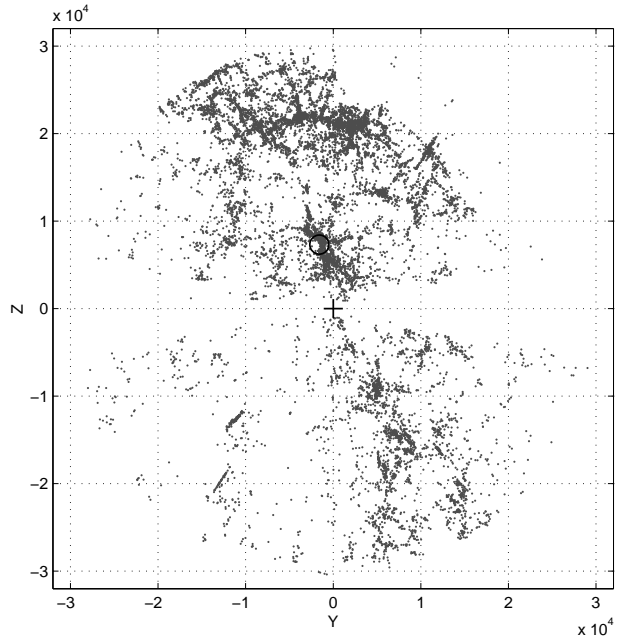} 
\caption{The galaxy distribution in a 1600~km/s thick slice of constant $X$ passing through the SC.  The SC is at the ``O", galactic north is upwards, and our galaxy is about 10400~km/s behind the central ``+".  See Fig. \ref{scm} for the slice position.}
\label{nedx}
\end{figure}

\begin{figure} 
\includegraphics[scale=0.6]{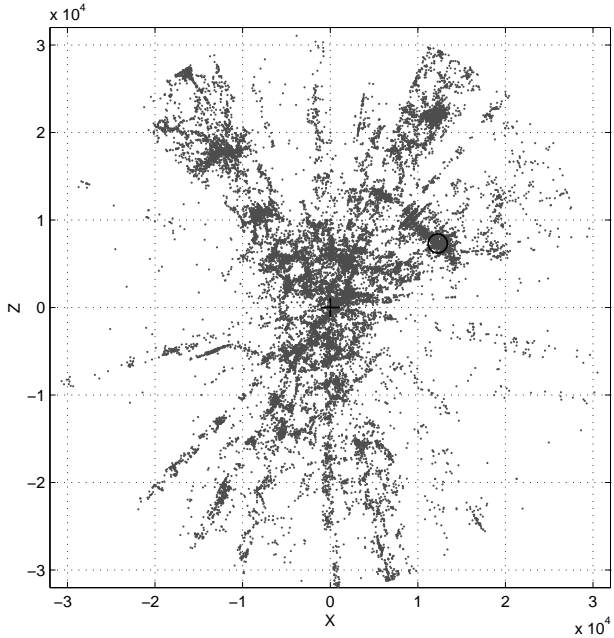} 
\caption{The galaxy distribution in a 1600~km/s thick slice of constant $Y$ passing through the SC.  The SC is at the ``O", our galaxy is at the central ``+", and up is galactic north.  See Fig. \ref{scm} for the slice position.}
\label{nedy}
\end{figure}

\begin{figure} 
\includegraphics[scale=0.6]{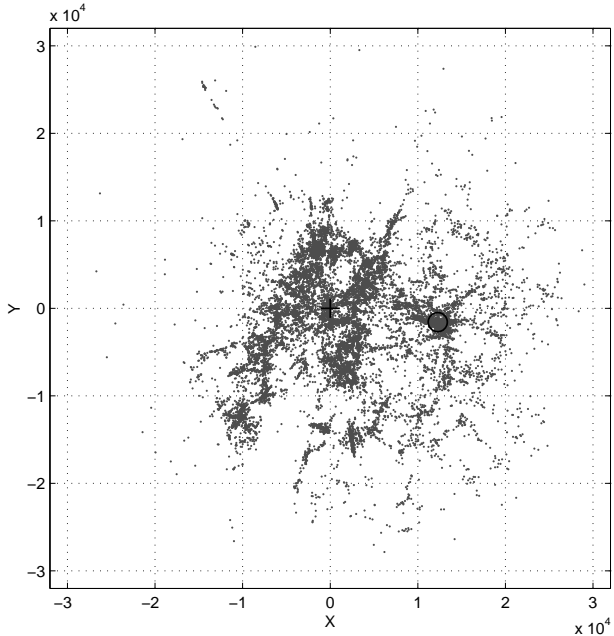} 
\caption{The galaxy distribution in a 1600~km/s thick slice of constant $Z$ passing through the SC.  The SC is at the ``O", our galaxy is about 7200~km/s behind the central ``+", and the $X$ direction corresponds to galactic longitude $\ell = 320^\circ$.  See Fig. \ref{scm} for the slice position.}
\label{nedz}
\end{figure}

 Fig. \ref{scm} presents the schematic position of the Local Group (LG), the Great Attractor (GA) and the Shapley Concentration (SC) in a coordinate system with the $X$ direction in the galactic plane pointing towards the GA, and the $Z$ axis pointing to galactic north. Figs. 2 -- 4 depict the Shapley concentration with surrounding galaxies, in perpendicular slices 1600~km/s thick and 32000~km/s square. Fig. \ref{nedx} shows the galaxy distribution in a constant $X$ slice through the SC, Fig. \ref{nedy} in a slice of constant $Y$, and Fig. \ref{nedz} a slice of constant $Z$.  The galaxy data were taken from the NASA/IPAC Extragalactic Database%
 \footnote{http://nedwww.ipac.caltech.edu/}%
 . As one can see, there are still regions in which there are hardly any observations, while many strips covered by very detailed galaxy surveys are evident. Furthermore, the largest amount of data are from our immediate neighborhood, deceptively suggesting that we are living in highly overdense region.  Fig. \ref{nedy} is the only slice that includes the local group, and it shows clearly how well measured the nearby regions are, how patchy the coverage is further out, and how the north is better surveyed than the south.  Unfortunately, individual surveys do not cover all the region of interest to us.  Thus there is significant uncerainty in the galaxy distribution around the SC, but we feel there is enough data to make an interesting model.  However Figs. 2--4 do show that the SC is a high density region, surround by regions of lower density, with a number of overdensities in various directions further out, such as the GA and the Hercules supercluster (visible at the top of Fig. 2). 

This paper provides a new approach to modeling these structures.  The innovation is that we model them as a one connected system, and we intend the models to be generally consistent with the astronomical data about the Shapley Concentration and the surrounding regions.  We create a relativistic non-linear model of the Great Attractor and the Shapley Concentration using the Lema\^itre--Tolman (LT) metric, which assumes spherical symmetry about some centre.  We view the SC as a major attractive centre, so we place the spherical origin at the SC, and we consider a sphere that includes the LG.  

\begin{figure} 
\includegraphics[scale=0.45]{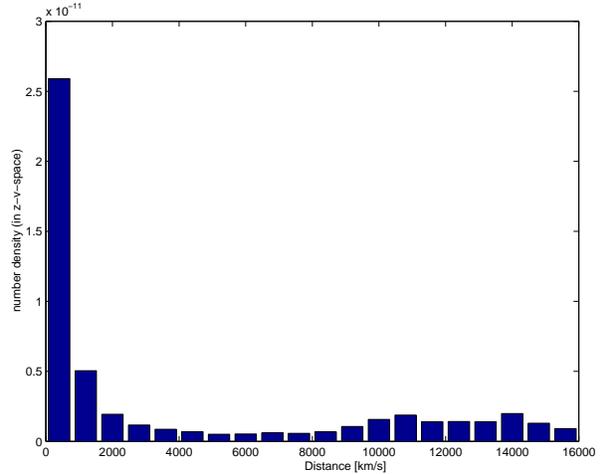}
\caption{The number density of galaxies in shells centred on the Shapley concentration (arbitrary units), normalised to compensate for the general decrease in the galaxy detection rate with distance from earth, as well as the differences in survey coverage in the north and south galactic regions and the zone of avoidance.  The SC is on the left, and the LG at the right.}
\label{Nsd}
\end{figure}

Although Fig. \ref{nedy} shows the core of the SC is not of spherical shape, this is at least partly an artefact of the non-uniform observational coverage in this region.  On large distance scales however, as can be seen in Figs. \ref{nedx} and \ref{nedz}, this assumption is not a bad first approximation to the galaxy data.  We further justify the assumption of an overdense sphere around the SC at a radius of about 14000~km/s, with a less dense region between, as follows.  We take the geocentric NED data, calculate the number of galaxies in a sequence of redshift/velocity bins, and fit it with a smooth polynomial.  The deviation of this polynomial from a constant value represents the decreasing fraction of galaxies that are detected with distance. 
 We also find the average galaxy density in the north galactic cap ($10^\circ < b \leq 90^\circ$), in the zone of avoidance ($-10^\circ \leq b \leq 10^\circ$), and in the south galactic cap ($-90^\circ \leq b < -10^\circ$).  Next we calculate the numbers of galaxies in a set of redshift/velocity shells centered on the SC, but we normalise the galaxy numbers at each point using the polynomial and the average densities in the 3 galactic lattitude zones.  The resulting histogram is shown in Fig. \ref{Nsd}.  The density maximum at the radius of the GA is quite distinct. 
  Also, in practice the galaxies are in a number of concentrations distributed around the sphere, such at the GA itself, so the actual density in each concentration is higher.  We intend that our model will be most accurate for regions around the SC, the GA and the LG, and it will naturally be considerably less accurate for regions further out than the SC, because of the scarcity of data there.

We aim to address the following questions: What is the mass and density distribution around the Shapley Concentration and the Great Attractor?  What is the peculiar velocity and density distribution in the region between the GA and SC?  Is there a backside infall towards the GA?  Does the SC have a significant impact on the motion of the LG?  Finally we are interested in investigating the influence of the cosmological constant on the dynamics of the local Universe.  Additional constraints on our model are derived from considering its evolution.  Since it is based on an exact solution of the Einstein field equations, we are able to trace the evolution backward (and forwards) in time (without any limit on the size of the density fluctuations), and we require the model to evolve from small initial fluctuations, preferably without forming shell crossings.

The structure of this paper is as follows: in Sec. \ref{obsc} we present the observational data about this region, in Sec. \ref{ltb} the Lema\^itre--Tolman model is presented, and Sec. \ref{res} presents the results of our calculations.

\section{Observational constraints}\label{obsc}

Direct astronomical observations of galaxies provide us with their positions on the sky and their redshifts.  With various methods of distance estimation (such as Tully--Fisher for spiral galaxies, $D-\sigma$ for elliptical galaxies or the SBF method) it is possible to convert redshifts into the peculiar velocities of the measured galaxies.  Once this is done, the masses of clusters can be estimated.  Masses can also be estimated from the flux of HI emission or X-rays.  Below we summarise the astronomical data about the Shapley Concentration and the surrounding regions, with which we wish our model to be consistent.

\begin{figure} 
\includegraphics[scale=0.7]{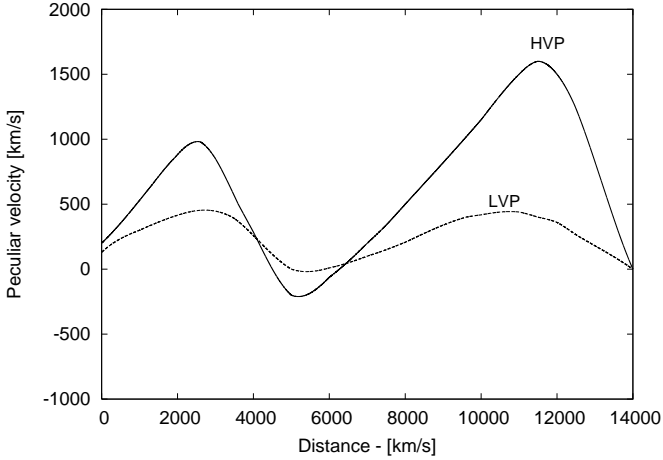} 
\caption{The main present day peculiar velocity profiles used in our models.  Distance is measured from the LG, with the SC on the right, for easy comparison with published estimates.}
\label{veld}
\end{figure}

\begin{figure} 
\includegraphics[scale=0.7]{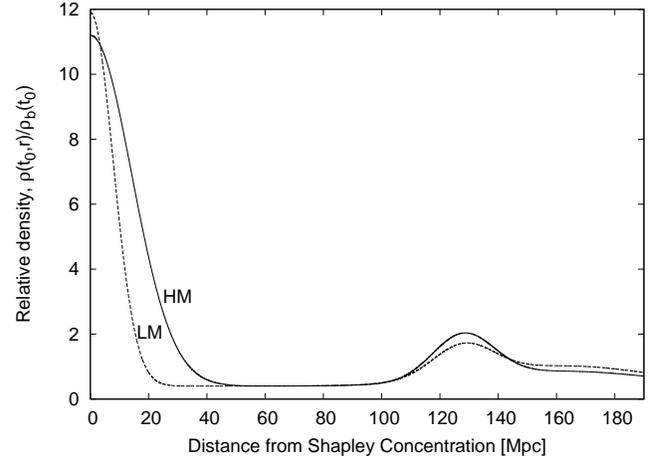} 
\caption{The main present day density distributions for the Shapley Concentration and surrounding regions, as used in our models.  Here distance is measured from the SC, with the radius of the LG on the right.}
\label{dend}
\end{figure}

\begin{enumerate}
\item
Peculiar velocities:
\begin{itemize}
\item
The motion of the Local Group with respect to the CMB is $627$~km/s (Tonry et al. 2000).
The contributions to this motion from the gravitational influence of the GA and the SC have been estimated by many astronomers.  Hoffman et al (2001) estimated that motion of the LG towards the GA is in the range $100$ to $200$~km/s.  Smith et al. (2000) estimated that $50 \pm 10 \%$ of the LG's motion is generated by the GA and the SC. 
The X-ray observations (Kocevski, Mullis \& Ebeling 2004) imply that the SC has a significant impact on the LG's motion.
\item
Tonry et al. (2000) conclude that the SC does not have a significant influence on the LG's motion, but their results show that the LG's motion towards the GA is in the range $200$ to $300$ km s$^{-1}$, and the infall velocity at a distance of $17$ Mpc from the center of the GA is $\sim 900$ km s$^{-1}$.
\item
Observations of clusters in the GA and the SC region imply a smaller variation in the peculiar velocity profile (Lucey, Radburn-Smith \& Hudson 2005).  According to their research, most clusters associated with the GA, have positive peculiar velocities (away from us) of $\sim 400$ km s$^{-1}$; clusters that are immediately beyond the GA have small peculiar velocities and do not show any evidence of backside infall; peculiar velocities of clusters in the region between GA and SC are of very small positive amplitude. 
\item
Observations of a few SnIa beyond $100$ Mpc show high positive values of the peculiar velocity, around the $1500$ km s$^{-1}$, however the error bars are large as well (Lucey, Radburn-Smith \& Hudson 2005).
\end{itemize}

The data above imply two possible profiles for the peculiar velocity
 along the line through the SC --- a high amplitude velocity profile (HVP) and a low amplitue velocity profile (LVP).  Fig. \ref{veld} presents both these profiles. 

\item
Mass distribution:
\begin{itemize}
\item
We expect the current mass distribution to have evolved from small initial fluctuations that existed at last scattering.  Therefore, given the dust equation of state, the total mass of our system today is the same as the mass of the same comoving sphere at last scattering, in other words the average density over the whole region is the same as the density of the background cosmology.
\item
Bardelli et al. (2000) estimated that the SC has mass $5.25 \times 10^{15}$~M$_\odot$ within a sphere of radius $14.03$~Mpc, and mass $8.25 \times 10^{15}$~M$_\odot$ within $19.6$ Mpc, based on the observed excess of galaxies.
\item
X-ray observations (Filippis, Schindler \& Erben 2005) suggest lower mass values; specifically $2.35 \times 10^{15}$~M$_\odot$ within a radius of $13.6$ Mpc, and $2.88 \times 10^{15}$~M$_\odot$ within $16.7$.  However, these estimates are based on flux observations, and so the contribution from the matter between the clusters and in the outer regions of the clusters is neglected.  Therefore these estimates can be treated only as lower limits.
\item
The mass of the GA is hard to determine, since it is  in the Zone of Avoidance, and estimates vary from $2 \times 10^{15}$~M$_\odot$ (Staveley--Smith 2000) to $3.75 \times 10^{16}$~M$_\odot$ (Lynden--Bell et al. 1988).
\item
The density contrast in the GA lies between $\delta \approx 0.73$  (Tonry et al. 2000) and $\delta \approx 1.1$ (Kolatt, Dekel \& Lahav 1995).
\item
Although the region between the GA and the SC is not well known, it can be seen from Figs. \ref{nedx} -- \ref{Nsd} that the SC is surrounded by low density regions.  The density contrast in one of them --- the Bootes void --- is in the range $\delta \approx -0.84$ to $-0.66$ (Day et al. 1990).
\end{itemize}

Astronomical data suggest several density profiles.  Let us consider two of them.  Namely, HM with high mass values for the SC and the GA, and LM with lower masses for both.  These profiles are presented in Fig. \ref{dend}.  Both these profiles include a low density region between the SC and the GA.

\end{enumerate}
 
\section{The Lema\^{i}tre--Tolman model}\label{ltb}

The Lema\^{\i}tre-Tolman (LT) model is the spherically symmetric solution of Einstein's equations with a dust source (Lema\^{\i}tre 1933, Tolman 1934).  In comoving and synchronous coordinates, the metric is:   
\begin{equation}   
{\rm d}s^2 =  c^2{\rm d}t^2 - \frac{R'^2(r,t)}{1 + 2 E(r)}\ {\rm   
d}r^2 - R^2(t,r) {\rm d} \Omega^2, \label{ds2}   
\end{equation}   
where $ {\rm d} \Omega^2 = {\rm d}\theta^2 + \sin^2 \theta {\rm d}\phi^2$,    
$R' = \partial R / \partial r$, and $E(r)$ is an arbitrary function of $r$. Because of the signature   
$(+, -, -, -)$, this function must obey $E(r) \ge - \frac{1}{2}.$   
   
The Einstein equations can be reduced to the following:   
\begin{equation}\label{den}   
\kappa \rho c^2 = \frac{2M'}{R^2 R'},   
\end{equation}   
\begin{equation}\label{vel}   
\frac{1}{c^2} \dot{R}^2 = 2E(r) + \frac{2M(r)}{R} + \frac{1}{3} \Lambda R^2,   
\end{equation}   
where $\dot{R} = \partial R / \partial t$, $M(r)$ is another arbitrary function, $\kappa = 8 \pi G / c^4$, and we take $V(t, r) = \dot{R}(t, r)$ as the velocity of a worldline.
   
Equation (\ref{vel}) can be solved by a simple integration:       
\begin{equation}\label{evo}   
\int\limits_0^R\frac{ {\rm d} \tilde{R}}{\sqrt{2E(r) + \frac{2M}{\tilde{R}} + \frac{1}{3}\Lambda \tilde{R}^2}} = c    
\left(t- t_B(r)\right),    
\end{equation}   
where the arbitrary function $t_B(r)$ appears as a function of integration.  This means that the Big Bang is not simultaneous for all worldlines, as in the Friedmann models, but occurs at different times for different coordinate distances from the origin.   
   
Thus, the evolution of the LT model is determined by three arbitrary functions: $E(r)$, $M(r)$ and $t_B(r)$. The metric and all the formulae are covariant under arbitrary (but monotonic) coordinate transformations of the form $r = f(r')$, and by use of such a transformation any one of these functions can be given many different forms.  Therefore the {\em physical} initial data for the evolution of the LT model consist of two arbitrary functions.

Apart from the Big Bang singularity that occurs when $R = 0$ along any given constant $r$ worldline (i.e. at $t = t_B(r)$), there are also shell crossings at $R' = 0$, $M' \ne 0$ where the density becomes infinite.  Shell crossings can be avoided by setting the initial conditions appropriately (Hellaby \& Lake 1985).
   
\subsection{The model setup}\label{proc}

We model a spherical inhomogeneous region centred on the SC, with a comoving LT radius $r_{max}$ that includes the local group.  We require that this region matches smoothly, in the Darmois sense, to a reasonable ``background" Robertson-Walker (RW) model at this outer radius.
 
The chosen background model is  the RW model with density 
 \begin{equation}
   \rho_b = 0.27 \times \rho_{cr} = 0.27 \frac{3H_0^2}{8 \pi G}.
   \label{rbdf}
 \end{equation}
 where the Hubble constant is $H_0 =72$ km $^{-1}$ Mpc$^{-1}$.  When models with a cosmological constant are considered, the value of $\Lambda$ corresponds to $\Omega_{\Lambda} = 0.73$, where $\Omega_{\Lambda} = (1/3)  ( c^2 \Lambda/H_0^2)$.

For the matching of the LT interior to the RW background, if $S_0$ and $\rho_0$ are the present day scale factor and density, $r_b$ is the standard RW radial coordinate, and $r_{bm}$ its value at the matching surface, then $2E(r_{max}) = - k r_{bm}^2$, $M(r_{max}) = \kappa \rho_0 (S_0 r_{bm})^3 /6$, and $t_B(r_{max}) = 0$ (i.e. we set $t = 0$ at the bang in the homogeneous background.)  Thus, the relative density inside the LT region is $\rho(t, r)/\rho_b(t)$, and the peculiar velocity of a worldline
 {\em relative to the SC} is 
$V(t, r) - V_b(t, r) = \dot{R}(t, r) - r \dot{S}_0(t)$.  However, for easy comparison with 
published peculiar velocity profiles, we plot the peculiar velocity {\em relative to the 
earth}, which, between the LG and the SC, is the negative of the foregoing, i.e. 
$r \dot{S}_0(t) - \dot{R}(t, r)$.  

For most of the following models, the sphere radius corresponds to the distance between the Shapley Concentration and the Local Group, $r_{max} = r_{LG}$ (about 14500~km/s $\approx$ 205~Mpc today), but in our last investigation it is slightly larger.

Methods for extracting the arbitrary functions that specify a LT model from given data such as density or velocity profiles have been given in Krasi\'nski \& Hellaby 2002, 2004a, 2004b; Bolejko, Krasi\'nski \& Hellaby 2005; Hellaby \& Krasi\'nski 2006.  Specifically, a given initial density profile $\rho(t_1, r)$ at time $t_1$ plus a given final density profile $\rho(t_2, r)$ at time $t_2$ fully determine a particular LT model.  In that work the coordinate choice $r = M$ was made.  Similarly, an initial and a final velocity profile $\dot{R}(t_1, r)$ and $\dot{R}(t_2, r)$, or one density profile and one velocity profie determine an LT model.  Amongst the other options presented, one is to specify a single density (or velocity) profile at a certain time and require a simultaneous bang time, which has the effect of removing decaying modes.  If one is seeking models that are free of shell crossings, this must be checked once the arbitrary functions have been calculated.

In this work, the procedure was adapted to the case of non-zero $\Lambda$, which means that the analytic evolution expressions of the zero-$\Lambda$ case had to be replaced by numerical determinations.

We started by choosing both a density profile and a velocity profile at the present day, solving for the arbitrary functions, and then looking at how the model evolved, especially the period from last scattering to the present day (section \ref{V&D}).  We found that this procedure typically gave severe shell crossings during the evolution, often not far in the past, which made the model useless.  

We then tried specifying only the present day velocity profile and requiring a simultaneous bang time $t_B =$~constant; and we also tried combining a present day density profile with a simultaneous bang time.  We found this last method was particularly good at providing models free of shell crossings (section \ref{SBT}).  

Having found a well-behaved model, it was then possible to tinker with it in order to adjust the present day profiles, whilst preserving a reasonable evolution (so the bang time was often not simultaneous in the final model).  The important point is that by considering how physically reasonable the evolution of the model is, we can put definite limits on the possible present day velocity profiles for a given present day density profile, and vice-versa.

\section{Results}\label{res}

\subsection{Present day velocity and density profiles}\label{V&D}

Astronomical observations based on different approaches provide us with different velocity and density profiles.  
In this section we aim to test the consistency of the various profiles with each other and with evolutionary scenarios.  
Although the LT model in principle allows us to match any initial velocity profile with any final density distribution, not all resulting evolutions are equally natural, and not all are free of shell crossings.  The velocity and density profiles used below are those of Figs. \ref{veld} and \ref{dend}.

\begin{figure} 
\includegraphics[scale=0.7]{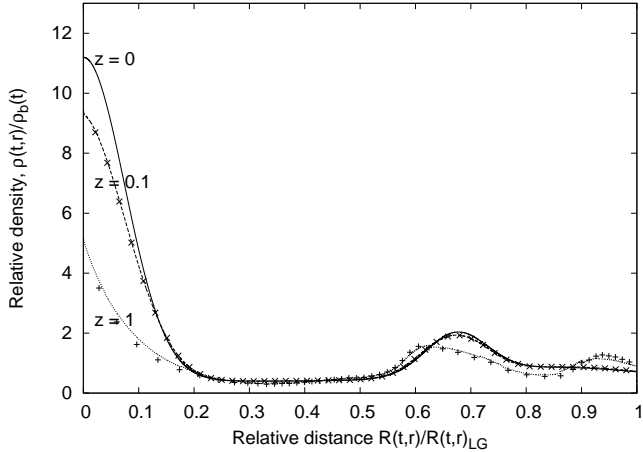} 
\caption{The density evolution of model 1.  The centre of the SC is at the left, the radius of the LG is at the right, and the radius of the GA is near relative distance 0.7.  Lines represent models with $\Lambda = 0$, points represent models with $\Omega_\Lambda = 0.73$.} 
\label{resmod1}
\end{figure}

\begin{itemize}
\item
Model 1 --- HVP + HM: \\
The evolution of this model is presented in Fig. \ref{resmod1}, which gives the relative density at each point, $\rho(t, r)/\rho_b(t)$.  The curves represent the evolution of the derived model without a cosmological constant, and the points represent the derived model evolution with $\Omega_\Lambda = 0.73$.  

Because the age of the Universe is different in various cosmological models, to compare models with and without $\Lambda$, we trace the evolution not in terms of time but in terms of redshift.  For example, at redshift $z = 1$ the age of the Universe is 5.9~Gy in a $\Lambda$-dust ($\Lambda$-CDM) model, and 4.6~Gy in a dust model without $\Lambda$.

\item
Model 2 --- LVP + HM: \\
The evolution results are given in Fig. \ref{resmod1}.  This model behaves differently from what might be expected.  The density at the center of the SC increases as we trace the evolution backwards.  This is due to a large central value of the bang time function, $t_B$, which at the origin is almost $3.5$ billion years more recent than elsewhere.

\item
Model 3 --- HVP + LM: \\
The evolution of this model is similar to that of model 1.

\item
Model 4 --- LVP + LM: \\
This model's evolution is quite like that of model 2.

\end{itemize}

The above results suggest that there is no significant difference between models with and without cosmological constant; in other words the value of $\Lambda$ is not important on this length scale even over cosmological timescales.  Additionally, models 2 and 4 have the SC growing towards the past, which is in contradiction with CMB observations.  Therefore there is a strong contradiction between the assumption of a low amplitude velocity proflie (LVP) and the requirement of a reasonable evolutionary scenario.

\begin{figure} 
\includegraphics[scale=0.7]{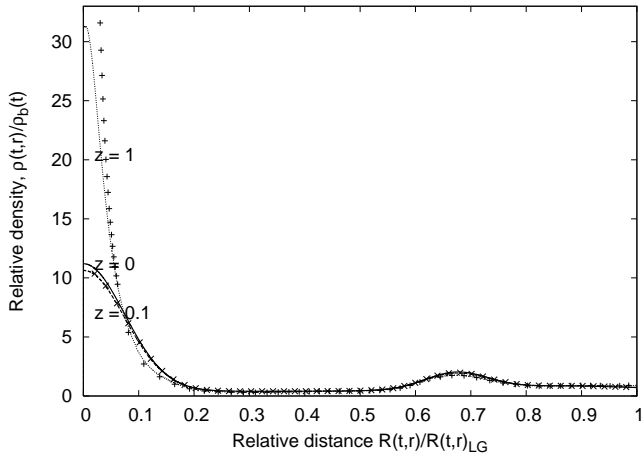} 
\caption{The density evolution of model 2.  Lines represent models with $\Lambda = 0$, points represent models with $\Omega_\Lambda = 0.73$.}
\label{resmod2}
\end{figure} 

\subsection{Simultaneous bang time}\label{SBT}

The results of the previous section imply that changes in the present day velocity profile have a large impact on the evolution of a model.  Moreover, the measurements of peculiar velocity have large errors, both observational and systematic (we are able to measure only the radial component of the velocity).   Thus, it is more sensible to obtain a calculated velocity profile from other considerations, and compare it with observations.

Measurements of the cosmic microwave background (CMB) imply that the Universe was very close to homogeneous at the time of last scattering, and as a consequence the variation in the bang time function cannot be larger than a few thousand years, which in comparison with the present age of the Universe is negligible.  Therefore, in this section the LT model is specified from a present day density distribution and the assumption $t_B = 0$.

\subsubsection{Peculiar velocities around the SC}

\begin{figure} 
\includegraphics[scale=0.7]{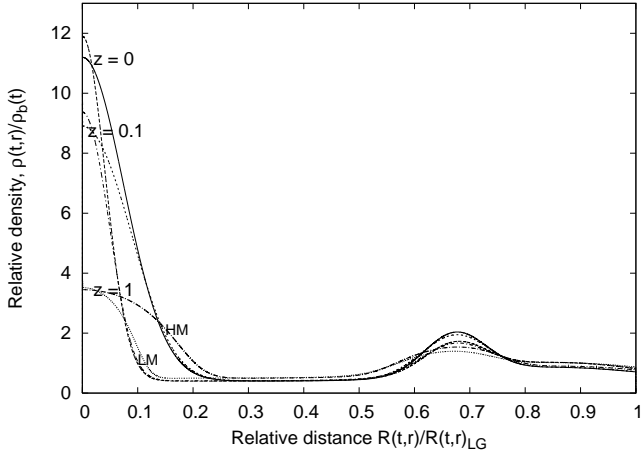} 
\caption{The evolution of models 5 (labelled ``HM") and 6 (labelled ``LM"), with $\Lambda = 0$.  The centre of the SC is at the left, the radius of the LG is at the right, and the radius of the GA is near relative distance 0.7.}
\label{resmod56}
\end{figure}

\begin{figure} 
\includegraphics[scale=0.7]{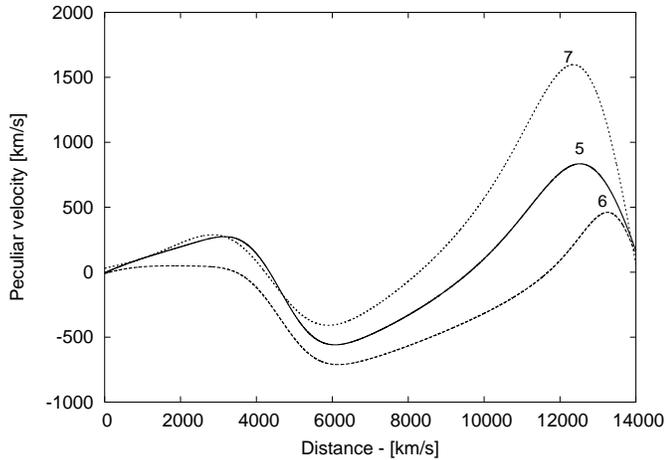} 
\caption{The present day peculiar velocity profiles used in models 5, 6 and 7.  The centre of the SC is at the right, the LG is at the left, and the GA is near 4000~km/s.}
\label{resmod567}
\end{figure}

\begin{itemize}
\item
Model 5 --- $t_B=0$ + HM: \\
The evolution of model 5 is presented in Fig. \ref{resmod56}.  For clarity, Fig. \ref{resmod56} only presents the evolution without a cosmological constant; evolution with a cosmological constant, as in the previous section, does not differ significantly from that without.  Since models with a simultaneous bang time are known to describe the evolution of growing modes only, the evolution of such models is pretty much as expected, and is free of shell crossings.  

The derived profile of the present day peculiar velocity is presented in Fig \ref{resmod567}.  As can be seen, the predicted amplitude of the peculiar velocity around the Shapley concentration is around 800 km s$^{-1}$, which is nicely within the range between the velocity profiles LVP and HVP.  However, this model predicts a significant backside infall onto the great attractor ($\sim - 590$ km/s in model the with $\Lambda = 0$ and $\sim - 620$ km/s in the model with $\Lambda \ne 0$), and very small values for the 
 maximum nearside infall onto the GA
 ($\approx 235$ km/s in the $\Lambda = 0$ model, and $\approx 250$ km/s in the $\Lambda \ne 0$ model).  Again the curves presented in Fig. \ref{resmod567} are for models with no cosmological constant, and differences in the peculiar velocity profiles between models with and without $\Lambda$ are nowhere larger than $ \sim 10 \%$.

\item
Model 6 --- $t_B=0$ + LM: \\
Fig. \ref{resmod56} shows the evolution of this model, and Fig. \ref{resmod567} shows the derived present day peculiar velocity profile.  Within a small distance from Shapley Concentration, the peculiar velocity is comparable with the LVP, but at all other distances the peculiar velocity drops well below observed values.  This result suggests that the low mass profile (LM) does not provide a realistic description of the density distribution around the Shapley Concentration.

\item 
Model 7 --- $t_B=0$ + $\sim$HVP: \\
This model aims to recover the high amplitude of peculiar velocities around the SC by adjusting the density profile.  The resulting density profile is obtained from HM by increasing the density in the SC region. 
To obtain the peculiar velocity maximum near the SC of $\approx 1600$~km/s, the mass of the SC must be almost doubled, i.e. within a radius of $14.03$~Mpc the mass should be $8.5 \times 10^{15}$~M$_\odot$, and within $19.6$~Mpc, $19.5 \times 10^{15}$~M$_\odot$.  Such a large mass contradicts the astronomical observations mentioned above.

Therefore, the results obtained in this and the previous section suggest a high mass profile (HM) around the Shapely Concentration, with peculiar velocity values of around 800~km/s.  However, all models so far considered, imply a significant backside infall onto the GA: $\sim$~-700~km/s (model 6), $\sim$~-600~km/s (model 5), or $\sim$~-400~km/s (model 7), which is not satisfactory.

\end{itemize}

\subsection{Backside infall}

In this section the issue of backside infall is investigated.  As stated, there is no observational evidence for any backside infall onto the GA.  Therefore, models 8 and 9, are intended to obtain a backside infall within the range $-100$~km/s to $+100$~km/s.  In both models we adopt a density distribution around the SC quite similar to the HM profile.

\begin{figure} 
\includegraphics[scale=0.7]{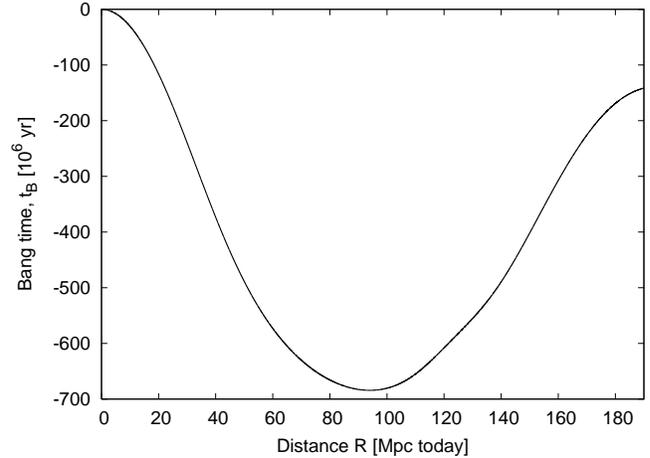} 
\caption{The bang time function $t_B$ obtained with model 10, against the present day separation of the corresponding worldlines.  The SC is on the left, and the LG is on the right.}
\label{tbm10}
\end{figure}

\begin{itemize}
\item
Model 8 --- $t_B=0$ + backside infall = $-100$~km/s: \\
To obtain a backside infall onto the GA of $-100$~km/s, the density {\em between} the SC and the GA must be around $\rho = 0.79 \rho_b$ in the model with $\Lambda = 0$ and around $\rho = 0.78 \rho_b$ in the model with $\Lambda \ne 0$, but otherwise similar to HM.

\item
Model 9 --- $t_B=0$ + backside infall = $+100$~km/s:

To obtain a backside infall of $+100$~km/s, the density between the SC and the GA must be about $\rho = 0.96 \rho_b$ with $\Lambda = 0$, and about $\rho = 0.95 \rho_b$ with $\Lambda \ne 0$, but otherwise similar to HM.

\item
Model 10 -- HM + backside infall = $0$~km/s:

This model aims to check the validity of the results of models 8 and 9, by investigating whether realistic models without backside infall really require the density between the SC and the GA to be quite close to background value, or lower intervening densities are also possible.

This model was specified by the HM density distribution and by a velocity distribution without backside infall
(similar to that for model 11, see Fig. \ref{lgm}).

{\em If} the bang time function $t_B(r)$ of this model had turned out to have only small variations (e.g. 1000 years), we would have found a realistic model without backside infall and with low density between the SC and the GA; one whose evolution after last scattering would be well behaved.  The bang time function we actually obtained from this model is presented in Fig. \ref{tbm10}.  The variation of $t_B$ between the SC and the GA is almost 700 million years.  Since this is far larger than the time of last scattering, it results in extensive shell crossings and large density variations at last scattering.

The above results show that the absence of backside infall does indeed require the density between the SC and the GA to be close to the background density.

\end{itemize}

Models 8, 9 and 10 show that the extent of backside infall onto the GA is very sensitive to the density between the SC and the GA.  If there is no backside infall, the density in this region must be $\rho \approx 0.85 \rho_b$.  Models without backside infall and with a void between the SC and the GA have strong variations  in the bang time function, which are not consistent with CMB observations.

\subsection{The motion of the Local Group}

So far all our models assumed that the mass inside the sphere of radius $r_{LG}$ is the same as the mass in a uniform sphere of the same size with a density equal to the background value.  This assumption, combined with the $t_B=0$ assumption, guarantees that there is no peculiar motion of the Local Group towards the GA.  
Now varying $M/E^{3/2}$ at $r_{LG}$ determines how much our worldline is gravitationally 
bound to the interior mass, while varying $t_B$ only affects how much of the evolution 
determined by $M/E^{3/2}$ has happened so far.  Thus we keep $t_B$ constant and manipulate $M$.  This induces a change of the mass within the sphere of radius $r_{LG}$, in comparison with the homogeneous background sphere.  Thus the matching to background happens somewhere further out. 

The assumed density distribution is the similar to HM, but instead of a void between the 
SC and the GA, it has density $0.9 \rho_b$, so it does not produce any backside infall.  
We then modified this density distribution, as explained below, so as to reproduce the 
LG's 200~km/s velocity toward the GA.

From eq. (\ref{vel}) it follows that to fit the value of 200~km/s for the LG's motion 
we have to increase $M(r_{LG})$.  This increase must be about $\Delta M \approx 95 \times 
10^{15}$~M$_{\odot}$ if $\Omega_\Lambda = 0.73$, and $\Delta M \approx 115 \times 
10^{15}$~M$_{\odot}$ if $\Omega_\Lambda = 0$.  The mass of the homogeneous background 
within that radius is $\approx 1115 \times 10^{15} M_{\odot}$ so the increase is around 
$\approx 10 \%$. 
 
This extra mass cannot be added to the SC, i.e. inside a sphere of radius 30~Mpc.  This 
is because the SC mass is far smaller, and such an increase would cause a larger than 
observed amplitude in the peculiar velocity profile.  Similarly, adding the mass in the 
low density between the SC and the GA would lead to an increased peculiar velocity in 
that region of more than $100$~km/s, as noted previously.  Therefore, the only possible 
way to increase the mass without inconsistency, is to add it at distances comparable with 
the GA's radius or larger.  Let us consider these possibilities:

\begin{itemize}
\item 
Model 11 --- increased GA mass: \\
A third possibility is to increase the density of shells at the radius of the GA.  
The density contrast at the GA's radius was increased from the initial $\delta = 1$ to 
$\delta = 1.4$, thus inducing a change of the Great Attractor mass%
 \footnote{To estimate this mass we take a roughly cubic blob out of the spherical density maximum near radius 130~Mpc from the SC.}
 from an initial $M \approx 4.7 \times 10^{15} M_{\odot}$ to $M \approx 5.9 \times 10^{15} 
M_{\odot}$.  The velocity distribution around the GA predicted by this model is in quite 
good agreement with the HVP (but only around the GA).

\item
Model 12 --- increased LG mass: \\
The last possibility is to increase the density of the region at the radius of the Local Group by $\Delta \delta \approx 0.1$.  The prediction of this model for the velocity distribution around the GA is also in quite good agreement with the LVP (but only around the GA).
\end{itemize}

\begin{figure} 
\includegraphics[scale=0.7]{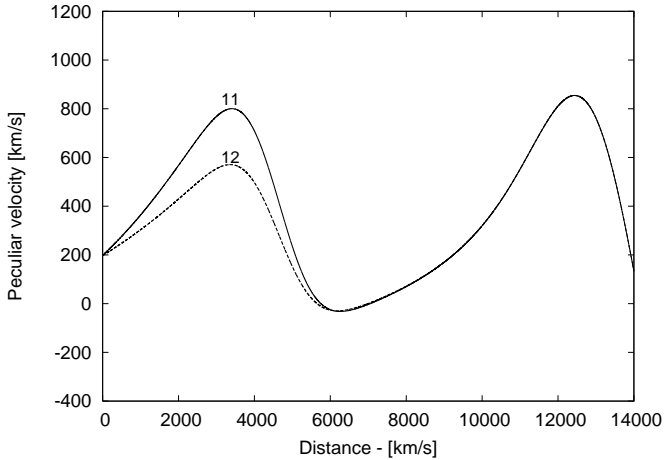} 
\caption{The peculiar velocity profiles that result from fitting the LG's motion towards GA, using models 11 and 12.  The curves are labelled by their model numbers.  The LG is on the left, and the SC is on the right.}
\label{lgm}
\end{figure}

The deduced peculiar velocities for these models are presented in Fig. \ref{lgm}, where for clarity, only results from the $\Lambda \ne 0$ model are plotted.

As can been seen the peculiar velocity around the GA is very sensitive to its mass.  A mass change of order $25 \%$ leads to a peculiar velocity change of about $50 \%$. So if we could extract from observation the exact value of the peculiar velocity profile around GA, we could estimate its mass to good precision.  However, due to the GA being in the Zone of Avoidance, this is a very difficult task.

\subsubsection{The SC's contribution to the LG motion}

The motion of the Local Group with respect to the CMB is around $600$ km s$^{-1}$.  The  cause of  this motion is still a riddle.  
The contribution to this motion from different superclusters and the Local Void have been systematical studied.  
However, there still remains the question of the Shapley concentration's contribution to this motion.  
For example, Tonry et al. (2000) suggest that the SC's effect is rather negligible.  On the other hand X-ray observations (Kocevski, Mullis \& Ebeling 2004) imply that the SC makes a significant contribution to the motion of the Local Group.

\begin{figure} 
\includegraphics[scale=0.7]{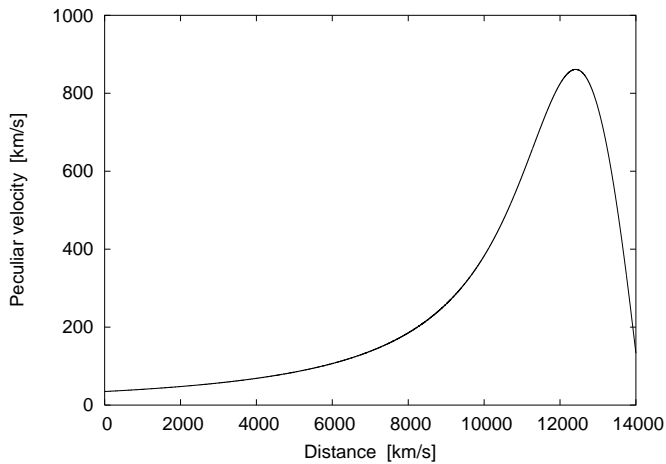} 
\caption{The peculiar velocity profile for a Universe consisting of the SC added to an otherwise uniform density.  The SC is at the right.}
\label{vlg2sc}
\end{figure}

To investigate this issue, we considered a model in which the density distribution is that of the Shapley Concentration in the HM profile out to where its relative density drops to 1 (around 30~Mpc), and then remains at the homogeneous background value beyond that.  The peculiar velocity thus obtained is presented in Fig. \ref{vlg2sc}.  It is apparent that the motion towards the SC at the distace of the LG is very low, approximately 35~km/s.

The explanation for this phenomenon can be illustrated by the following approximate calculation based on the LT evolution formula (\ref{vel}).  Let us consider a uniform gravitating sphere of radius $R$ and mass $M$, with a small extra mass $m$ (e.g. the SC) at the center, $m << M$.  The influence of this additional mass on the velocity of test particles at distance $R$ is estimated as follows:
\begin{eqnarray}
\frac{1}{c} \Delta V &=& \frac{1}{c} \left( V - V_b \right) =  \sqrt{2 E + \frac{2(M+m)}{R} + \frac{1}{3} \Lambda R^2} + \nonumber \\
&& -  \sqrt{2 E + \frac{2M}{R} + \frac{1}{3} \Lambda R^2} \nonumber \\
&=&  \sqrt{(2 E + \frac{2M}{R} + \frac{1}{3} \Lambda R^2) (1 + \frac{2m c^2 }{R \dot{R}^2})} + \nonumber \\
&& - \sqrt{2 E + \frac{2M}{R} + \frac{1}{3} \Lambda R^2} \nonumber \\
 &\approx& -\frac{m c}{R \dot{R}} \approx - \frac{ m c }{ R^2 H_0}.
\end{eqnarray}
In the above formula, a negative velocity means motion towards the center, hence the peculiar velocity at radius $R$ is towards the extra mass.  Converting the above relation to a ready-to-use formula, we obtain:
\begin{equation}
\Delta V = 4.3 \times 10^4 \frac{ \mathcal{M}}{ D^2 h},
\end{equation}
where $\mathcal{M}$ is the extra mass in units of $10^{15}$~M$_{\odot}$, $D$ is the distance in Mpc, $h$ is defined by $H_0 = 100\,h$~km/s/Mpc, and $\Delta V$ is the amount of infall velocity due to the central mass excess in km/s.  
Therefore, although the mass of the Shapley Concentration is large (${\mathcal M}\sim 10$), the distance to it is also large, ($D\sim 190$) and hence its contribution to the motion of the Local Group is small.

\section{Conclusions}

We have developed a relativistic approach to modeling the Great Attractor and the Shapley Concentration, with the best models being models 11 and 12.  The Great Attractor region has been the subject of many observational and theoretical studies.  
However, due to the Great Attractor being in the Zone of Avoidance, all of these studies had to make some assumptions.  The POTENT analysis (Bertschinger \& Dekel 1989; Dekel, Bertschinger \& Faber 1990a, 1990b; Kolatt, Dekel \& Lahav 1995) assumed Newtonian gravity and no vorticity.  Tonry's infall analysis (Tonry et al. 2000) assumed a spherical infall onto the Great Attractor, which implied the existence of backside infall.

Our approach instead assumes spherical symmetry around the Shapley Concentration.  
This assumption allows us to investigate whether or not a flow towards Great Attractor can be a part of a larger flow towards the Shapley Concentraton.
Although, our spherical assumption is only a first approximation, it is not seriously off, especially for the region this side of the SC.  Our research should be viewed as one of several complementary approaches to studying the Great Attractor.  The overlap of all these approaches should provide us with better understanding the Great Attractor region.

The conclusions of our study of the SC, GA and LG region are as follows:

\begin{enumerate}
\item
The peculiar velocity around the SC is $\sim 800$~km/s towards its centre.
\item
The lack of evidence for backside infall onto the GA implies that the density between the GA and the SC must be about $\delta \approx 0.9$.
\item
The mass of the GA is around $4 - 6 \times 10^{15}$ M$_\odot$.
\item
The SC's contribution to the LG peculiar motion is negligible.
\item
On the scales of this investigation, the value of the cosmological constant does not have a significant impact on the formation and evolution of structure.
\end{enumerate}

\section*{ACKNOWLEDGMENTS}
We would like to thank Patrick Woudt, Tony Fairall and Rene Kraan-Korteweg of UCT's Astronomy Department for their valuble comments and discusions concerning the Great Attractor, the Shapley Concentration, and the local galaxy distribution.  
KB is very grateful to CH and UCT's Department of Mathematics and Applied Mathematics, where most of this research was carried out, for their hospitality.  CH thanks the South African National Research Foundation for a grant.  An award from the Poland-South Africa Technical Cooperation Agreement is gratefully acknowledged.

\end{document}